\documentclass[preprint,
amssymb,floatfix,floats,aps,preprintnumbers,noshowpacs,
]{revtex4-1}
\usepackage{lipsum}
\usepackage{color,graphicx}
\usepackage{xcolor}
\usepackage{float}
\usepackage{amssymb}
\usepackage{ctable}

\newcommand{\cps}{Ce$_3$Pd$_{20}$Si$_6$}

\begin{document}
	
	\title{$^{105}$Pd NMR and NQR study of the cubic heavy fermion system \cps}
	
	\author{I.~Jakovac$^1$, M.~Horvati\'c$^2$, E.~F.~Schwier$^3$, A.~Prokofiev$^4$, S.~Paschen$^4$, H.~Mitamura$^5$, T.~Sakakibara$^5$, M.~S.~Grbi\'c$^{1,\dagger}$}
	
	\affiliation{$^{1}$Department of Physics, Faculty of Science, University of Zagreb, Bijeni\v {c}ka cesta 32, Zagreb HR 10000, Croatia}
		     \affiliation{$^2$Laboratoire National des Champs Magn\'etiques Intenses, LNCMI-CNRS (UPR3228), EMFL, Universit\'e Grenoble Alpes, UPS and INSA Toulouse, Bo\^{i}te Postale 166, 38042, Grenoble Cedex 9, France}
		    \affiliation{$^3$Hiroshima Synchrotron Radiation Center, Hiroshima University, Higashi-Hiroshima, Hiroshima 739-0046, Japan}
		    \affiliation{$^4$Institute of Solid State Physics, Vienna University of Technology, 1040 Vienna, Austria}
		    \affiliation{$^5$Institute for Solid State Physics, University of Tokyo, Kashiwa 277-8581, Japan}

	
	\begin{abstract}
	We report $^{105}$Pd NMR and NQR measurements on a single crystal of \cps, where antiferroquadrupolar and antiferromagnetic orders develop at low temperature. From the analysis of NQR and NMR spectra, we have determined the electric field gradient (EFG) tensors and the anisotropic Knight shift ($K$) components for both inequivalent Pd sites - Pd($32f$) and Pd($48h$). The observed EFG values are in excellent agreement with our state-of-the-art DFT calculations. The principal values of the quadrupolar coupling are $(20.37 \pm 0.02)$~MHz and $(5.45 \pm 0.02)$~MHz,  for the Pd($32f$) and Pd($48h$) site, respectively, which is large compared to the Larmor frequency defined by the gyromagnetic constant $\gamma = 1.94838$~MHz/T for $^{105}$Pd. Therefore, the complete knowledge of $K$ and the EFG tensors is crucial to establish the correspondence between NMR spectra and crystallographic sites, which is needed for a complete analysis of the magnetic structure, static spin susceptibility, and the spin-lattice relaxation rate data and a better understanding of the groundstate of \cps.
	\end{abstract}
\maketitle	
	%
	%
	%
	%
	%

\section{Introduction}
The ``caged'' cerium compound \cps\ ~\cite{Strydom_2006,PASCHEN200790} is of great interest since the discovery~\cite{Custers} that this Kondo system shows signatures of Kondo destruction quantum critical point (QCP)~\cite{SiKondo}. On decreasing temperature in zero magnetic field, the systems enters first an antiferroquadrupolar (AFQ) state~\cite{Goto2009,Portnichenko2016} at $T_Q= 0.45~$K and then an antiferromagnetic (AF) state~\cite{KITAGAWA199748,Mitamura2010} at $T_{AF}= 0.23~$K. In a magnetic field of $\approx0.8~$T, the AF state is continuously suppressed and a QCP emerges, while the AFQ state remains stable. Whereas initially it was considered that Kondo destruction quantum criticality requires low-dimensional spin fluctuations~\cite{SiKondo}, more recent considerations of the global phase diagram of antiferromagnetic heavy fermion compounds~\cite{SiRev2013} have clarified that, within an ordered phase (the AFQ phase in finite magnetic fields exhibits field-induced dipolar moments \cite{Portnichenko2016}), this is not required~\cite{Custers}. The AFQ phase is suppressed only at larger magnetic fields, near 2 T along [001] direction. Interestingly, also at this second QCP, Kondo destruction phenomena are observed~\cite{Martelli2019}, and the sequence of two QCPs was understood as the step-wise incorporation of the spin and the orbital components into the Kondo cloud at the two distinct QCPs ~\cite{Martelli2019}. The compound has also raised interest as it has one of the heaviest electron masses among the Kondo systems, with a low-temperature electronic specific heat coefficient~\cite{KITAGAWA1996} of $\gamma_{el} = 8$~J/mol$_{\rm{Ce}}$~K$^2$ that arises due to the proximity to the QCPs. \cps\ is unique as it has two inequivalent Ce sites and, as such may represent the case when two values of Kondo screening are present in the system,  theoretically studied by Benlagra et al.~\cite{Benlagra2011}. Its exotic phase diagram has been studied by various techniques ~\cite{Mitamura2010,Deen2010,Portnichenko2015,Portnichenko2016}, and a detailed neutron scattering investigations have shown that both the AFM and the AFQ phase are due to moments on the $8c$ site~\cite{Portnichenko2015, Portnichenko2016}. Thus, the question arises what is the fate of the 4$f$ electron on the second Ce site? So far, there are 
only two 
nuclear magnetic resonance (NMR) studies of a system demonstrating Kondo destruction phenomena, both done on YbRh$_2$Si$_2$~\cite{IshidaPRL,KambeNat}, which makes another NMR investigation of this phenomenon on a different system of great value. The NMR studies on YbRh$_2$Si$_2$ revealed the emergence of critical fluctuations at low temperatures and magnetic fields, in contrast to the Fermi liquid (FL) behaviour observed at higher fields. Knight shift ($K$) and spin-lattice relaxation rate ($1/T_1$) measurements were used to determine the uniform static spin susceptibility $\chi'(q=0)$ and $1/T_1$ divided by temperature ($1/T_1 T$), respectively, the latter characterizing the spin fluctuations by providing the \textbf{\textit{q}} averaged dynamical spin susceptibility $\chi''(q,\omega)$. If both quantities are temperature independent, a FL state is evidenced. Temperature dependence, on the other hand, informs on the nature of spin fluctuations at the QCP.
		
In this paper, we report measurements of the nuclear quadrupolar resonance (NQR) and NMR of $^{105}$Pd nuclei, along with a DFT analysis of the local electric field gradients (EFG) at the Pd sites. We determine the complete EFG tensor and the values of Knight shift components for all the Pd sites. The complete knowledge of these quantities will allow to determine the temperature dependence of the Knight shift and a full analysis of the $1/T_1$ data in future studies.
	
				\begin{figure}[t]
		\centering
		\includegraphics[width=0.5 \linewidth]{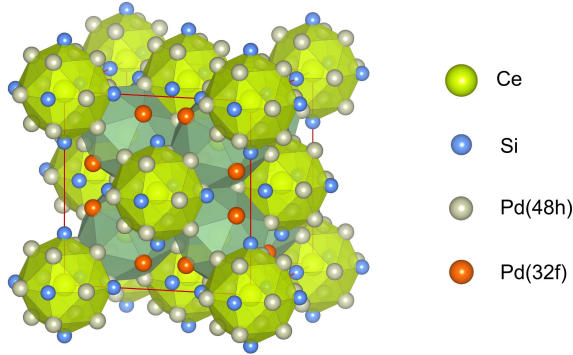}
		\caption{The crystal structure of \cps. The symmetry inequivalent Ce($4a$) and Ce(8$c$) positions are enclosed by light and dark green cages, respectively.}
		\label{fig:structure}
	\end{figure}
	
		\begin{figure}[b!]
	\centering
	\includegraphics[width=0.6\linewidth]{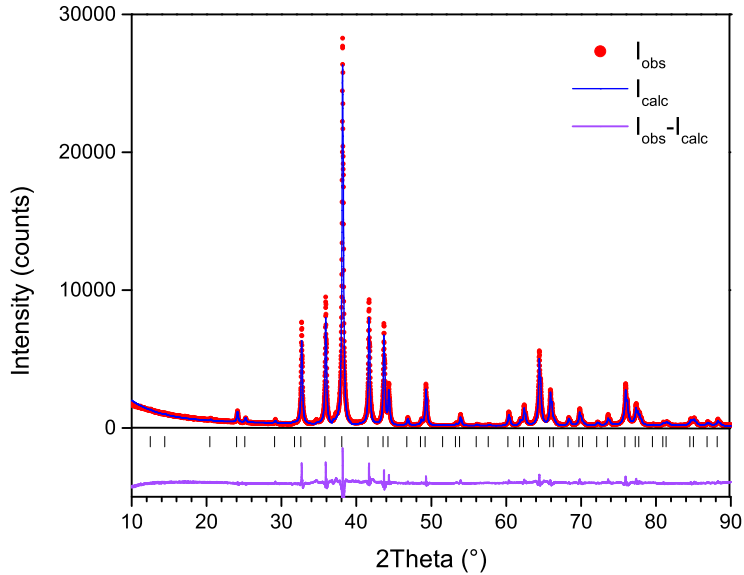}
	\caption{X-ray diffraction of powdered crystals of \cps\ (red dots) and the fit (blue line) for the cubic $Fm\bar3m$ space group with a lattice parameter $12.2749(2)$~\r{A} (upper panel), and a residual of the data with respect to the fit (lower panel). Black ticks correspond to the reflections
for the $Fm\bar3m$ structure.}
	\label{fig:XRD}
\end{figure}
	
	\section{Crystal structure and experimental methods}
	
	\cps\ crystallizes in the cubic $Fm\bar3m$ space group (Fig.~\ref{fig:structure}) with two crystallographically inequivalent cerium sites: Ce($4a$) and Ce($8c$) ~\cite{gribanov}. The unit cell comprises 4 formula units (116 atoms) with the lattice parameter $a = 12.161$ to $12.280$~\r{A}, depending on stoichiometry~\cite{gribanov,Prokofiev2009,PaschenProkofiev,Takeda1995}. The Ce atoms at the $4a$ Wyckoff position are surrounded by $12$ Pd and $6$ Si atoms, have an octahedral ($O_h$) point symmetry and form an FCC sublattice. The Ce atoms at the tetrahedrally symmetric ($T_d$) $8c$ position are caged by 16 Pd atoms and form a simple cubic structure nested in the FCC sublattice. The Pd atoms also occupy two inequivalent sites, namely Pd($48h$) and Pd($32f$). While Ce($4a$) is enclosed exclusively by Pd($48h$) atoms, Ce($8c$) has four neighbouring Pd($32f$) atoms, in addition to the shared Pd($48h$) atoms. Finally, the Si atoms occupy the 24e Wyckoff position.

	A single-crystalline Ce$_3$Pd$_{20}$Si$_6$ investigated in this work was grown from a slightly off-stoichiometric melt in a four-mirror furnace using the floating zone technique ~\cite{Prokofiev2009,PaschenProkofiev}. X-ray diffraction of powdered crystals indicates a single phase material with a= 12.2749(2) \r{A} (Fig.~\ref{fig:XRD}) and SEM/EDX investigation shows the composition to be 10.7 at.$\%$ Ce, 68.1 at.$\%$ Pd, and 21.2 at.$\%$ Si, in a good agreement with the stoichiometry of \cps. 
	 The sample was cut into a $4 \times 6$~mm$^2$ platelet, approximately 0.07~mm thick, with one crystallographic axis, taken to be [001], along the longer side. Such orientation is convenient for rotational NMR and NQR measurements. The sample was placed on a single-axis goniometer with this [001] axis aligned with the rotation axis. Therefore, the magnetic field was oriented in the [010]-[100] plane as the sample was rotated.
		
	NMR measurement were performed by varying the magnetic field intensity (magnetic field sweep), while the signal was measured at $f=26.7$~MHz and $20$~K. At this frequency we were able to observe the major part of the very broad NMR spectra (Fig.~\ref{fig:nmr}) in a magnet that can provide magnetic field values up to 15~T, and thus reconstruct the values of the local NMR parameters. At 20~K we expected the highest signal-to-noise ratio, while remaining in the paramagnetic state~\cite{new}. The resolution of the magnetic field sweep was $70$~mT. The sample was rotated from $+6^\circ$ to $-64^\circ$ degrees with respect to crystallographic [010] axis (goniometer angle $\alpha$). For signal acquisition, we used a Hahn echo sequence ($\pi/2 - \tau - \pi$) with a typical duration of the $\pi/2$ pulse of 4.8~$\mu$s and $\tau= 36~\mu$s. The measured spectra were fit with a multi-peak Gaussian curve.   
	
	    \section{Density functional theory calculations}
Density functional theory (DFT) calculations were performed using
the ELK code~\cite{elk_code,Lejaeghereaad3000}, which employs the all-electron full-potential linearised augmented-plane wave method. A $5 \times 5 \times 5$
$k$-mesh was used to sample the Brillouin zone. To account for the variation between different functionals, we used both the local density approximation (LDA)~\cite{LDA} and generalized gradient approximation (GGA)~\cite{GGA} including non-collinear fully relativistic spin-orbit coupling. The variation between LDA and GGA
is taken as error of the theory when comparing with experimental results.	
	\section{NMR spectra and quadrupole splitting}
	In general, the NMR Hamiltonian can be written as
		\begin{equation}
	\label{eq:ham}
	\mathcal{H} = \mathcal{H}_Q + \mathcal{H}_M,
	\end{equation}
	 where $\mathcal{H}_Q $ corresponds to the nuclear quadrupole interaction and $\mathcal{H}_M$ represents the Zeeman interaction. The quadrupolar interaction term arises due to the coupling of the nuclear quadrupolar moment to the local electric field gradient (EFG) and can be expanded as:
	
	\begin{equation}
	\label{eq:nqr}
	\mathcal{H}_Q = \frac{h\nu_Q}{6}\left[3I_z^2-I^2+\frac{\eta(I_+^2 + I_-^2)}{2}\right],
	\end{equation}
	\noindent with the characteristic quadrupole coupling constant defined as:
	\begin{equation}
	\label{eq:nuq}
	\nu_Q = \frac{3eQ}{2I(2I-1)}V_{zz}.
	\end{equation}
	\noindent Here, $e$ represents the elementary charge, $I$ and $Q$ stand for nuclear spin and quadrupole moment, respectively, $I_z$ and $I_+$($I_-$) stand for the standard $z$ projection and raising (lowering) operators of the nuclear spin, respectively, while $V_{zz}$ denotes the principal value of the local EFG tensor. $h$ stands for the Planck constant. In the absence of magnetic field the resonant frequencies are determined by separation of the eigenvalues of $\mathcal{H}_Q$. For $^{105}$Pd, $I=5/2$ and $Q=6.60 \cdot 10^{-29}$ m$^2$. The asymmetry parameter $\eta$ is defined by $\eta = (V_{yy}-V_{xx})/V_{zz}$ and takes values in
the interval $[0,1]$, due to constraints imposed by the
Laplace equation ($\nabla^2V = 0$) and the standard
convention $|V_{xx}| \le |V_{yy}|\le|V_{zz}|$. The symmetry of the
quadrupolar Hamiltonian implies double degenerate
energy levels, and for $I = 5/2$ we thus expect to
observe two NQR lines. For the axial ($\eta = 0$) case, pure Zeeman states provide a trivial solution of the
Hamiltonian (\ref{eq:nqr}), and the two frequencies are $f_1 = \nu_Q$ and $f_2 = 2\nu_Q$, corresponding to the transitions
$\pm\frac{1}{2} \rightarrow \pm\frac{3}{2}$  and $\pm\frac{3}{2} \rightarrow \pm\frac{5}{2}$, respectively.
Anisotropy ($\eta \ne 0$) introduces mixed states, and the
two transition frequencies are~\cite{MladenThesis,abragam}:
	
	\begin{eqnarray}
	f_1 =  \frac{2 \ \nu_Q \ r}{\sqrt{3}}\sin(\phi/3), \\
	f_2 = \nu_Q \ r \left(\cos(\phi/3)-\frac{1}{\sqrt{3}}\sin(\phi/3)\right),
	\end{eqnarray}
	where $\phi$ and $r$ are defined as:
	\begin{equation}
	\phi =  \arccos \left(10 \frac{1-\eta^2}{r^3}\right),
		\end{equation}
		\begin{equation}
	r = \sqrt{7 (1+\eta^2/3)}.
	\label{eq:nqr2}
	\end{equation}
	As a function of $\eta$, the $f_2/f_1$ ratio monotonically
decreases~\cite{MladenThesis,Chudo_2010} from 2 at $\eta= 0$ to 1 at $\eta= 1$, where $f_1 = f_2 = 2\sqrt{7} \nu_Q / 3$. The observed two NQR line positions (frequencies) thus directly define the values of
both $\nu_Q$ and $\eta$.

	The degeneracy of $\mathcal{H}_Q$ is lifted through the Zeeman interaction when the system is perturbed by a magnetic field:
		\begin{equation}
	\label{eq:nmr}
	\mathcal{H}_M = -\hbar \gamma \textbf{H} (1+\hat{K}) \textbf{I},
	\end{equation}

	\noindent where $\gamma$ is the gyromagnetic ratio of the probed nuclei ($1.94838$~MHz/T for $^{105}$Pd), $\textbf{H}$ for the magnetic field vector, $\textbf{I}$ for the total spin and $\hat{K}$ stands for the Knight shift tensor. With $K_x$, $K_y$ and $K_z$, we mark its components in the directions of the principal axes of the local EFG tensor. Both coordinate systems of principal axes are equivalent due to the well defined local symmetries at both Pd sites, and the overall cubic symmetry of the lattice. We discuss the orientation of the principal axes for each Pd site in the next section. The shape of the NMR spectrum is defined by the direction of the magnetic field ($\vartheta_{EFG}$, $\varphi_{EFG}$) with respect to the principal axes of the EFG tensor~\cite{abragam}:
	
    \begin{equation}
	\label{eq:ham_nqr}
 \textbf{H}= |\textbf{H}| \pmatrix{
 	\sin(\vartheta_{EFG}) \cos(\varphi_{EFG}) \cr \sin(\vartheta_{EFG}) \sin(\varphi_{EFG})\cr \cos(\vartheta_{EFG})}.
	\end{equation}
\begin{figure}[b!]
	\centering
	\includegraphics[width=0.4\linewidth]{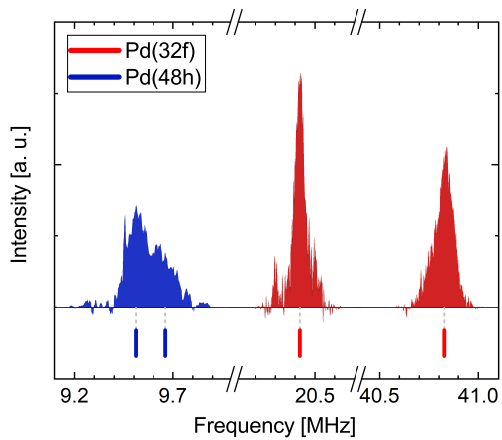}
	\caption{NQR spectra of $^{105}$Pd nuclei at $T=20$~K. The two sets of split NQR lines reveal different symmetries of the two Pd sites.}
	\label{fig:nqr}
\end{figure}	
	
	\begin{table}[h]
		\caption{Quadrupole coupling constant ($\nu_Q$) and asymmetry parameters ($\eta$) for the sites Pd($32f$) and Pd($48h$), as determined by DFT calculations and by NQR measurements, by use of (\ref{eq:nuq}).}
		\centering
		\begin{tabular}{cccccc}
			\specialrule{.1em}{.05em}{.05em}  
			site & $\nu_Q$ [MHz] & $\eta$ \\
			\hline
			Pd($48h$)$^{\rm{GGA}}$   & $7.05 $ & $0.68$ \\
			Pd($48h$)$^{\rm{LDA}}$   & $6.64 $ & $0.86$ \\
			Pd($48h$)$^{\rm{exp.}}$ & $5.46 \pm 0.05$ & $0.98 \pm 0.02$ \\
			\hline
			Pd($32f$)$^{\rm{GGA}}$  & $22.73 $ & $0$ \\
			Pd($32f$)$^{\rm{LDA}}$  & $21.78 $ & $0$ \\
			Pd($32f$)$^{\rm{exp.}}$ & $20.40 \pm 0.03$ & $0.03 \pm 0.03$ \\ 
			
			\specialrule{.1em}{.05em}{.05em} 
		\end{tabular}
	\label{tab:dft}
	\end{table}
	
		\begin{figure*}[t!]
		\centering
		\includegraphics[width=0.9\linewidth]{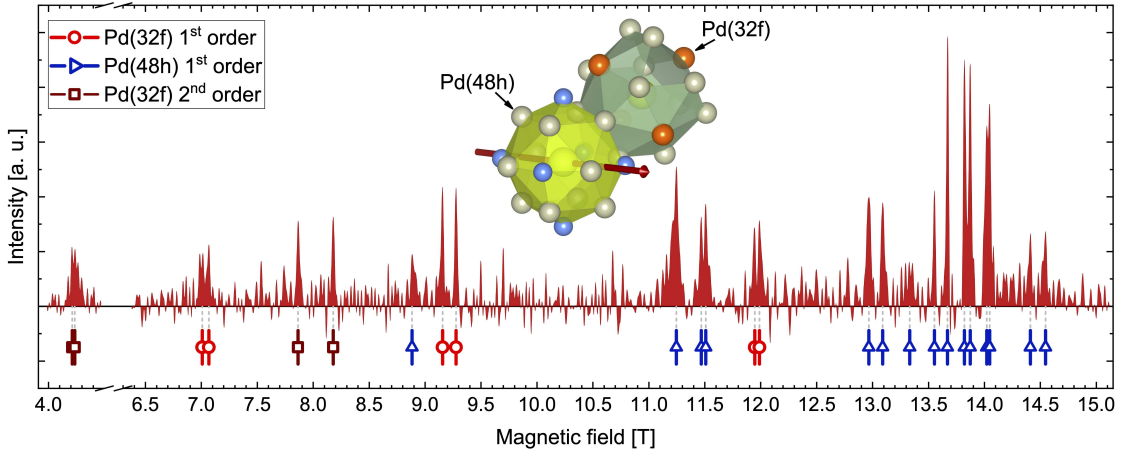}
		\caption{Field-sweep NMR spectra at $20$~K for the $\mathbf{H}$ vector (red arrow) at an angle $\alpha = 33.7^\circ$ to the $[0 1 0]$ crystallographic axis. A strong perturbation by an external magnetic field splits the NQR lines, revealing first ($\Delta m = \pm 1$) and  second ($\Delta m = \pm 2$) order transitions~\cite{comment}.}
		\label{fig:nmr}
	\end{figure*}

	\section{Results}
	\subsection{DFT calculation}
We employed DFT to calculate the EFG tensor. Similar calculations were previously used to analyze photoelectron spectroscopy results of this system~\cite{DFT}. After rotation to the system of local principle axes, the LDA calculations give $V_{xx} =4.684\cdot10^{21}$~V/m$^2$, $V_{yy} =4.684\cdot10^{21}$~V/m$^2$ and $V_{zz} =-9.368\cdot10^{21}$~V/m$^2$ for the Pd($32f$) site, while for the Pd($48h$) site the values are $V_{xx} =0.020\cdot10^{21}$~V/m$^2$, $V_{yy} =2.652\cdot10^{21}$~V/m$^2$ and $V_{zz} =-2.853\cdot10^{21}$~V/m$^2$.

The orientation and sign of the calculated EFG tensors respect the local symmetries, which is discussed in detail in section~\ref{subNQR}. These values were converted to the parameters $\nu_Q$ and $\eta$, and the results are shown in Table~\ref{tab:dft}. Therefore, from DFT we expect one pair of lines at a higher frequency, in the vicinity of 22 and 44~MHz, and a pair of lines closer to each other at a lower frequency of about 12~MHz.

 \subsection{NQR measurements}
 \label{subNQR}
	The NQR measurements of $^{105}$Pd, taken at $20$~K, are shown in Fig.~\ref{fig:nqr}. The lines at 20.41~MHz and 40.83~MHz can be described by a Gaussian function; they present the resonant frequency ratio of approximately 2:1 and thus they correspond to the axially symmetric  ($\eta=0$) position Pd($32f$). At the low-frequency end, close to $\approx9.6~$MHz, there are two almost completely superimposed spectral lines which correspond to the asymmetric site Pd($48h$) with $\eta \approx 1$. 

	The values of $\nu_Q$ and $\eta$ determined from the measurements are listed  in Table~\ref{tab:dft}. The agreement between data and DFT is surprisingly good: while for Pd($32f$) the difference between the predicted and the measured value of $\nu_Q$ is less than $10\%$, for Pd($48h$) this difference is  $\simeq 20\%$. This is a rather astonishing agreement, if we consider that in Ce-based compounds the existing spatial localization of the correlated $f$ electrons is {\it{not}} very well described and hence the calculated electric fields ought to be affected throughout the unit cell. 
	Measured values of the quadrupolar coupling constants for both sites are comparable to the values determined on $^{105}$Pd nuclei in other materials shown in Table~\ref{tab:nuQval}. Only one of these studies compares experiment with theory ~\cite{Chudo_2010}: for $\nu_Q$ a difference of $30\% - 55\%$ was reported for CePd$_5$Al$_2$.

	\begin{table}[t]
	\caption{List of quadrupolar coupling constants of $^{105}$Pd in other systems. The asterik (``*") marks two sets of possible values that were determined from the acquired data.
	}
	\centering
	\begin{tabular}{ccc}
		\specialrule{.1em}{.05em}{.05em}
		compound & reference  & $\nu_Q (^{105}$Pd$)$ (MHz)  \\ 
		\hline
		NpPd$_5$Al$_2$ & \cite{Chudo_2010} &  12.04 (Pd(1)), 35.34 (Pd(2)) \\ 
		
		CePd$_5$Al$_2$ & \cite{Chudo_2010}  &  11.31 (Pd(1)), 39.96 (Pd(2))   \\
		
		UPd$_2$Al$_3$ & \cite{Matsuda1997,MATSUDA1998447,Kohori,MATSUDA1999640}  & 5.317 \\
		
		UPd$_3$ & \cite{FUJIMOTO2007746} & 22.5* (Pd(1)), 36.7* (Pd(2)) \\
		
		UPd$_3$ & \cite{FUJIMOTO2007746} & 28.1* (Pd(1)), 34.9* (Pd(2)) \\
		\specialrule{.1em}{.05em}{.05em}
	\end{tabular}
	\label{tab:nuQval}
\end{table}

\subsection{NMR measurements}
\begin{figure*}[t!]
		\centering
		\includegraphics[width=0.95\linewidth]{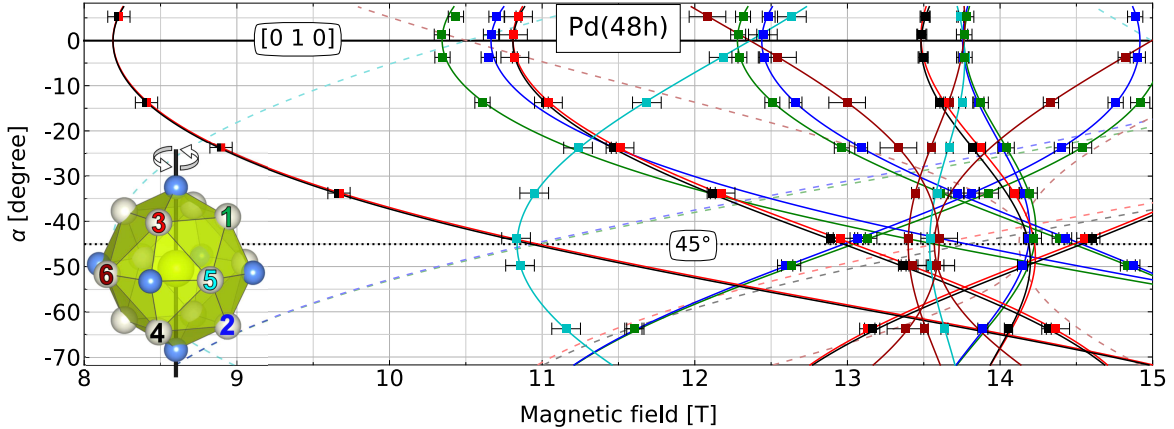}
		\caption{The angular dependencies of NMR transitions observed at $26.7$~MHz and $20$~K for six inequivalent positions resulting from the Pd($48h$) site (inset) when an external magnetic field brakes the symmetry. The sample was rotated about the [001] axis and the data were fit by (\ref{eq:ham}). Full lines show the fitted angle dependence obtained by exact diagonalization of the Hamiltonian: the angle dependence of transitions that were not observed are indicated by dashed lines. The fit parameters are summarized in Table \ref{tab:nmr}. The color code defines the correspondence between the Pd positions (inset) and the NMR line positions and fits. }
		\label{fig:pd48hfit}
	\end{figure*}

	To determine the components of the shift tensor $\hat{K}$ we made a magnetic field sweep acquisition of NMR signals at the frequency of 26.7~MHz, and observed how the line positions depend on the orientation of the applied field. A typical spectrum is shown in Fig.~\ref{fig:nmr}. Due to the large quadrupolar coupling, and a low $\gamma$ value of $^{105}$Pd, even the strong magnetic field of 15 T only perturbs the zero-field Hamiltonian, as we do not reach the Zeeman limit where $\mathcal{H}_M \gg \mathcal{H}_Q$. Hence, to analyse the data the total system's Hamiltonian (\ref{eq:ham}) has to be numerically diagonalized. The multistep procedure involves defining the NMR line positions from the spectra for each angle of the goniometer and connecting the data to the site-specific ($\theta_{EFG}$,~$\phi_{EFG}$) angle pairs, defined by respective crystallographic positions that enter expression (\ref{eq:nmr}). The rotational dependence of the NMR line positions are shown in Fig.~\ref{fig:pd48hfit} and ~\ref{fig:pd32ffit}.
	
	A global numeric fit of all the parameters for each crystallographic site was performed in order to determine the parameters common for the entire dataset, such as the exact direction of the rotation axis and the initial direction of the magnetic field vector. The rotation axis is confirmed to coincide with the crystallographic [001] axis within the experimental error. For the Pd($48h$) site all the parameters are varied, while for the Pd($32f$) site's spectral data the fitting was constrained by the $\eta = 0$ and $K_x = K_y$ conditions to comply with the specific (axial) site symmetry. The respective parameter uncertainties are determined by the shape of the square error ($\chi^2$) minimum, that is its second-order derivatives (Hessian matrix). The diagonal elements of the covariance matrix, which is the inverse of the Hessian matrix, are identified as the parameter variances. The results of the fits to the NMR spectral data are shown in Table \ref{tab:nmr}. Because of the larger number of points  measured in NMR than in NQR, the $\nu_Q$ and $\eta$ values have higher statistical precision than the former. The parameter error bars determined by NQR are dominated by the conservative approach of taking spectral line width as the uncertainty of the characteristic frequencies $f_{1,2}$.

				\begin{figure}[h]
	\centering
	\includegraphics[width=0.5\linewidth]{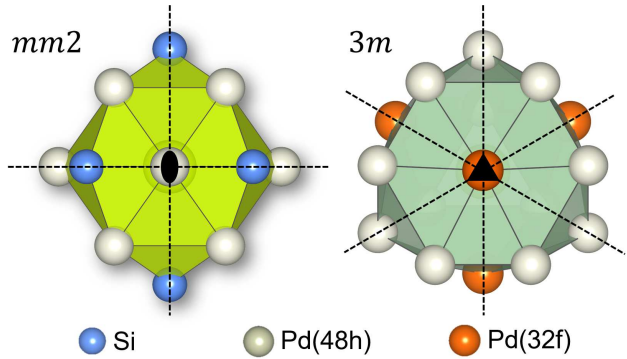}
	\caption{Part of the crystal structure surrounding Ce($4a$) (left) and Ce($8c$) (right) site, respectively. Each cage is rotated such that the point symmetry of the two Pd sites is highlighted. Dashed lines represent the mirror planes while the black oval and triangle mark the 2-fold and 3-fold crystallographic rotation axis. In both cases the direction of $V_{zz}$, the principal value of the EFG tensor, coincides with these axes. For Pd($48h$) these are the face diagonals of the (cubic) unit cell, and for Pd($32f$) the space diagonals. $V_{xx}$ and $V_{yy}$ are normal to mirror planes for the Pd($48h$) site and undefined for Pd($32f$) due to the axial symmetry of the EFG tensor.}
	\label{fig:symmetry}
\end{figure}
\subsection{Local site symmetries}
    A thorough understanding of local site symmetry is essential for spectral data interpretation, and we will now discus it for each Pd site.
    
	\subsection*{Pd($48h$) site}
	Pd($48h$) site corresponds to the 12 equivalent positions for Pd that surround Ce($4a$) site (Fig.~\ref{fig:symmetry}) which have mm2 ($C_{2v}$) point symmetry, identified by the 2-fold rotation axis and two perpendicular mirror planes. Therefore, the principal axes of the EFG tensor should point along these directions. From DFT calculations it follows that the axis of the principal value $V_{zz}$ points along the rotation axis, i.e. in the crystallographic directions coinciding with face diagonals of the unit cell ($[1 1 0]$, $[1 0 1]$, $[0 1 1]$,...), depending on the exact position of the Pd($48h$) atom. The vectors normal to characteristic mirror planes define the other two principal axes. The calculated value of the asymmetry parameter $\eta \geq 0$ reflects the fact that indeed, as expected from the crystal structure, the associated EFG tensor does not possess axial symmetry. Fitting the spectra for Pd($48h$), shown in Fig.~\ref{fig:nqr}, by (\ref{eq:nqr}-\ref{eq:nqr2}) suggests a high asymmetry of the observed position with $\eta = 0.98 \pm 0.02$ and $\nu_Q = (5.46 \pm 0.05$)~MHz. The EFG eigenvalue V$_{xx}$ almost vanishes along the direction of the nearest Si atoms, i.e., in the directions $[1 \overline{1} 0]$, $[\overline{1} 0 1]$, $[0 \overline{1} 1]$,... This is not well reproduced by the DFT calculations, similar to the case of the Cu(1) site in YBa$_2$Cu$_3$O$_7$ ~\cite{Pennington89,Schwarz1990,IK1998}. To learn more about the origin of this feature we also performed a point-charge calculation of the EFG tensor of the ``lattice" contribution, i.e. without taking into account the hybridization of the Pd and Si orbitals. The results clearly indicate that the lattice contribution is negligible in the resulting EFG tensor and that the effect of local polarization of $p$-orbitals in Si~\cite{DFT} is the dominant cause of the V$_{xx} \approx 0$ value.

	When the site symmetry is broken by an external magnetic field, six different sets of NMR lines from six different Pd positions resulting from the Pd($48h$) site can be identified (Fig.~\ref{fig:pd48hfit} inset); the other 6 positions of the Pd($48h$) site remain centrally symmetric with respect to the caged Ce($4a$) atom, and thus provide identical spectra. We note that in Fig.~\ref{fig:pd48hfit} we have shown angle dependences of both the central ($\pm1/2 \leftrightarrow \mp 1/2$) transition (marked with circles) and the low-field first-order satellites (square markers). Hyperfine coupling of $^{105}$Pd to nearby ions will result in a Knight shift expressed by $K_x$, $K_y$ and $K_z$ parameters in (\ref{eq:nmr}). Because of symmetry, the Knight shift tensor will have the same eigenbase as the EFG tensor. Furthermore, we expect $K_x$, $K_y$ and $K_z$ to take different values due to lack of axial site symmetry. Respective values of $K_x$, $K_y$, $K_z$, $\nu_Q$ and $\eta$ are calculated by fitting the total system Hamiltonian (\ref{eq:ham}) to the entire set of measured points and the results are presented in Table \ref{tab:nmr}. We find an excellent agreement with the $\nu_Q$ and $\eta$ values determined from NQR. 

	\begin{figure}[t!]
	\centering
	\includegraphics[width=0.55\linewidth]{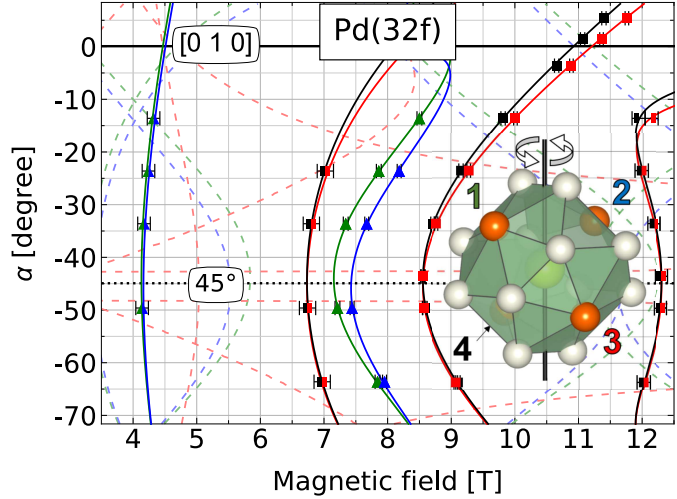}
	\caption{Angular dependence of the first ($\Box$) and the second~\cite{comment} ($\triangle$) order NMR transitions observed at $26.7$~MHz for the Pd($32f$) crystallographic site (inset). The sample was rotated around the [001] axis with magnetic field in the [100]-[010] plane. Measured data were fitted by (\ref{eq:ham}) and the spectral lines assigned to the respective Pd positions (color-coding). The 45$^\circ$ line emphasizes  the high-symmetry axis. The error bars correspond to the spectral line width (Gaussian). We observe two different types of rotation behaviour for 1-2 and 3-4 positions, respectively. Full lines mark the fitted angle dependence obtained by exact diagonalization of the Hamiltonian. The angle dependence of the unobserved transitions is given by dashed lines.}
	\label{fig:pd32ffit}
\end{figure}	
	
	\subsection*{Pd($32f$) site}
	The Pd($32f$) site describes 4 equivalent Pd positions in the cage around each Ce($8c$) atom. Each position has 3m ($D_{3d}$) point symmetry characterized by a 3-fold rotation axis and 3 mirror planes, as shown in Fig.~\ref{fig:symmetry}. The rotational axis then defines the axis of the EFG tensor principal value $V_{zz}$ which coincides with space diagonals of the unit cell ($[1 1 1],[\bar{1} 1 1],[1 \bar{1} 1],[1 1 \bar{1}]$). Due to the axial symmetry of the Pd($32f$) site, the asymmetry parameter $\eta$ vanishes, and the principal directions of the $V_{xx}$ and $V_{zz}$ values can be chosen freely in the plane perpendicular to $V_{zz}$ direction.

	In general, in an external magnetic field, the four inequivalent positions resulting from the Pd($32f$) site will produce different sets of spectral lines. However, due to the high symmetry of the Pd($32f$) site and a vanishing $\eta$, a rotation around the [001] axis with magnetic field perpendicular to the axis of rotation will produce only two sets of lines. Namely, it can be shown that pairs of Pd($32f$) atoms below and above the plane defined by the rotation axis will always make the same angle with the $\mathbf{H}$ vector direction. However, in the spectra shown in Fig.~\ref{fig:pd32ffit} there is a small splitting between the theoretically coinciding lines, due to imperfect sample orientation. It should be noted that Fig.~\ref{fig:pd32ffit} shows the central transitions (circles), and the first and second~\cite{comment} order satellites (squares and triangles, respectively). While the two sets of red and black lines correspond to the two first-order NMR satellites of Pd($32f$) positions 3 and 4, the two sets of blue and green lines correspond to the two second-order NMR satellites of Pd($32f$) positions 1 and~2.
	
	The same logic as for the Pd($48h$) site implies that $K_z$ points in the same direction as $V_{zz}$, and that the other two principal values, $K_x$ and $K_y$, are necessarily equal. The result of the fitting procedure performed as earlier are summarized in Table~\ref{tab:nmr}. The only additional constraint that was set here is that $\eta =0$. Again, we find a good agreement between the $\nu_Q$ values determined here with those found from NQR.

	\begin{table}[t]
	\caption{List of principal values of the Knight shift tensors, quadrupole coupling constants ($\nu_Q$) and asymmetry parameters $\eta$ at $20$~K for the two Pd sites. The principal axes of the Knight shift tensors are imposed by symmetry constraints.
	}
	\centering
	\begin{tabular}{ccc}
	\specialrule{.1em}{.05em}{.05em}
	parameter & Pd($48h$)  & Pd($32f$)  \\ 
	\hline
	$K_z$ [$\%$] & $0.40 \pm 0.05$ &  $0.19 \pm  0.09$ \\ 
	
	$K_y$ [$\%$] & $0.80 \pm 0.03$   & $-(1.4 \pm 0.04)$  \\
	
	$K_x$ [$\%$] & $-(1.51 \pm 0.04)$ & $-(1.4 \pm 0.04)$\\
	
	$\nu_Q$ [MHz] & $5.45 \pm 0.02$ & $20.37 \pm 0.02$ \\
	
	$\eta$ & $0.99 \pm 0.01$  & 0\\
	\specialrule{.1em}{.05em}{.05em}
\end{tabular}
\label{tab:nmr}
\end{table}

	\section{Conclusion}
	We have measured the $^{105}$Pd NMR and NQR spectra in a single crystalline sample of \cps\ and compared the results to DFT calculations. Each NMR/NQR signal has been assigned to the respective Pd positions that form cages around the corresponding Ce ions. By doing this we have characterized both Pd sites of the structure, as well as the inequivalent positions resulting from them in an applied magnetic field, and determined the components of its EFG and Knight shift tensor. We find a very good agreement between theoretical calculations and experiment.
	
	The complete correspondence between NMR spectra and crystallographic sites is important for future analyses of the magnetic structure, static spin susceptibility and also for a proper treatment of the spin-lattice relaxation data. Therefore, the results provide a basis for further investigations of the AFQ and AF orders, properties of the critical fluctuations and quantum critical phenomena that develop at low temperatures at two distinct magnetic fields~\cite{Martelli2019}.

	\section{Acknowledgments}
	M.S.G. and I.J. acknowledge the support of Croatian Science Foundation (HRZZ) under the project IP-2018-01-2970, the Unity Through Knowledge Fund (UKF Grant No. 20/15) and the support of project CeNIKS co-financed by the Croatian Government and the European Union through the European Regional Development Fund - Competitiveness and Cohesion Operational Programme (Grant No. KK.01.1.1.02.0013). A.P. and S.P acknowledge funding from the Austrian Science Fund (projects P29296 and P29279) and the European Union's Horizon 2020 Research and Innovation Programme under Grant Agreement No. EMP-824109. M.S.G. and I.J. thank Ivan Kup\v ci\'c for discussions and point-charge model calculations. 
	

\begin{thebibliography}{38}
\bibitem [{new()}]{new}%
	\BibitemOpen
	\href@noop {} {}\bibinfo {note} {The Kondo temperature is estimated to be
		$\approx 1~$K, so that at $\approx 10~$K we expect deviations in the Knight
		shift might occur due to the band hybridization effects.}\BibitemShut {Stop}%
\end{thebibliography}
	\bibliographystyle{plain}

\end{document}